\def\year{2025}\relax

\documentclass[letterpaper]{article} 
\usepackage{aaai25}  
\usepackage{times}  
\usepackage{helvet}  
\usepackage{courier}  
\usepackage[hyphens]{url}  
\usepackage{graphicx} 
\usepackage{natbib}  
\usepackage{caption} 
\DeclareCaptionStyle{ruled}{labelfont=normalfont,labelsep=colon,strut=off} 
\usepackage{subcaption}
\usepackage{longtable}
\usepackage{multirow}
\usepackage{booktabs}
\newcommand{\change}[1]{{\color{black}#1}}
\DeclareCaptionStyle{ruled}{labelfont=normalfont,labelsep=colon,strut=off} 

\usepackage{multirow}
\usepackage{algorithm}
\usepackage{algorithmic}
\usepackage{float}
\usepackage{xcolor}

\usepackage{newfloat}
\usepackage{listings}
\floatstyle{ruled}
\newfloat{listing}{tb}{lst}{}
\floatname{listing}{Listing}
\DeclareCaptionStyle{ruled}{labelfont=normalfont,labelsep=colon,strut=off}

\title{In Times of Crisis: An Exploratory Study of Media and Political Discourse on YouTube During the 2024 French Elections}
\author{
Vera Sosnovik, Caroline Violot, Mathias Humbert
}
\affiliations{
\textsuperscript{\rm}University of Lausanne\\
\{vera.sosnovik, caroline.violot, mathias.humbert\}@unil.ch
}

\begin{document}

\maketitle

\begin{abstract}
YouTube has emerged as a major platform for political communication and news dissemination, particularly during high-stakes electoral periods. In the context of the 2024 European Parliament and French legislative elections, this study investigates how political actors and news media used YouTube to shape public discourse. We analyze over 100,000 video transcripts and metadata from 73 French YouTube channels operated by national news outlets, local media, and political figures. To identify the key themes emphasized during the campaign period, we applied a semi-automated method that combined large language models with clustering and manual review. The results reveal distinct thematic patterns across the political spectrum and media types, with right-leaning news outlets focusing on topics like immigration, while left-leaning emphasized protest and media freedom. Themes generating the most audience engagement, measured by comment-to-view ratios, were most often the most polarizing ones. In contrast, less polarizing themes such as video games and nature showed higher approval, reflected in like-to-view ratios. We also observed a general tendency across all media types to portray political figures in neutral or critical terms rather than favorable ones. 

\end{abstract}


\section{Introduction}



YouTube has become an increasingly important platform for news consumption, particularly among younger audiences who turn away from traditional media. According to the 2024 Reuters Institute Digital News Report, there is a clear global shift toward video-based content, with platforms like YouTube gaining traction as news and political commentary sources~\cite{reuters}. Its accessibility, free use, and algorithm-driven personalization make it especially attractive to users seeking real-time information and a wider range of perspectives than those typically offered by mainstream broadcast outlets. With over two billion active users globally each month, YouTube serves not only as a major social media platform but also as a key source of news and information. Moreover, YouTube offers political parties and politicians a powerful platform to communicate directly with the public, bypassing the editorial gatekeeping of traditional broadcast media. This autonomy allows them to craft and control their messaging, tailor content to specific audiences, and respond rapidly to unfolding events. As a result, YouTube has become a strategic tool for promoting campaign narratives and shaping political discourse on its terms~\cite{30channels,ukpolit}. YouTube’s popularity, accessibility, rich metadata, and mix of political and media actors make it a powerful platform for analyzing and comparing evolving patterns of political communication and news media discourse.


While previous research has examined political discourse during elections across various social media platforms and countries~\cite{oliveira2018politicians, wu2021cross, covidfat, 10.1145/3665450, 10.1145/3543507.3583875, caro}, studies focusing specifically on political and news media discourse on YouTube during election periods remain scarce. In this context, the French 2024 elections present a unique and timely case study, shaped by political crisis and the rise of the far right. The European Parliament election, held on 9 June 2024, triggered a major political upheaval following the far-right Rassemblement National’s best-ever result~\cite{brookings}, surpassing President Macron’s Renaissance party. In response, Macron dissolved the National Assembly and called snap legislative elections. This period of political instability offers a valuable opportunity to examine how discourse unfolded across French news media and political communication on YouTube.


In this paper, we focus on the detection and analysis of themes emphasized by politicians and news media during the election period. Understanding these themes is crucial, as they reflect the strategic priorities of political actors and the editorial choices of media outlets in shaping public discourse. During elections, the issues that dominate political messaging and media coverage can significantly influence voter attention, frame the terms of debate, and ultimately impact electoral outcomes. By identifying which themes were highlighted and how they were presented, we gain insight into the dynamics of political communication and the role of media in framing democratic participation during a time of heightened political tension.

For this study, we analyzed YouTube channels operated by national and local news media, political parties, and politicians who participated in the 2024 European and French legislative elections. To capture both the buildup and immediate aftermath of these events, we analyzed content published between 1 March 2023 and 15 July 2024. Using the YouTube API to collect channel and video metadata, and the Transcript API to extract video transcripts, we compiled a dataset of over 100,000 video transcripts and related metadata from 73 selected channels. 

A central methodological challenge in analyzing political discourse during the 2024 French elections is identifying which themes to capture and whether existing frameworks are suitable for this context. In this study, we chose not to rely on predefined codebooks such as those from the Comparative Agenda Project (CAP) \cite{Baumgartner-2006} or the Comparative Manifesto Project (CMP) \cite{CMP}, despite their wide use in political content analysis. These frameworks are designed for long-term agenda tracking and may miss the breadth and specificity of themes that surfaced in news coverage and political messaging during the 2024 election period. Predefined categories also risk overlooking emergent or context-specific issues that were particularly salient at the time. 

To extract themes from the transcripts, we employed a multi-step, semi-automated approach. First, we used a large language model (Mistral's NeMo) to generate topic labels for each transcript~\cite{nemo}. The number of topics requested was proportionate to the transcript's length, ensuring appropriate granularity. These topics were then vectorized and clustered using the k-means algorithm to identify recurring themes across the dataset. The resulting clusters were manually reviewed and refined to merge semantically similar themes and improve interpretability. 

Our analysis of national news media channels on YouTube reveals clear thematic divides across the political spectrum. Left-leaning outlets emphasized \textit{News Media \& Freedom of Expression} and \textit{Protests}, largely absent from centrist and right-leaning coverage. Centrist channels focused on \textit{Geopolitics \& Conflicts}, while right-leaning media highlighted \textit{Immigration \& Refugees} and \textit{Health \& Health Care}. Some themes were highly polarizing: \textit{Religion \& Beliefs} were especially divisive on left-leaning channels, while \textit{Antisemitism} stood out on centrist and right-wing outlets. Notably, right-leaning news consistently drew more comments, indicating higher engagement or more reactive audiences. 

Analysis of politicians’ and parties’ YouTube channels reveals differences in thematic priorities. Right-wing politicians primarily addressed \textit{News Media \& Freedom of Expression}, while far-right actors focused more on \textit{Protests} theme. These themes were also frequently covered by left-leaning news media. In contrast, centrist and left-wing politicians devoted greater attention to \textit{Climate Change \& Environmental problems}.


\section{Data Collection}\label{data_collect}


We focused on three types of information sources. First, we included official YouTube channels of politicians and political parties who participated in either the European or French legislative elections to capture direct political messaging during the campaign period. Second, we selected national French news channels to analyze how mainstream media covered key political developments. Finally, we incorporated local French news sources to understand the issues and narratives emphasized at the regional level.

We compiled a list of parties and politicians participating in the 2024 European and French legislative elections~\cite{europ_el, leg_el} and manually searched for their official YouTube channels. For the European elections, we focused on party leaders and representatives whose parties secured seats in the European Parliament. For the legislative elections, we selected prominent participants, including major party leaders and high-profile candidates from across the political spectrum. In both cases, we began by checking official websites for links to YouTube channels; if none were available, we conducted manual searches using the individuals’ names directly on YouTube. A channel was included in our dataset if it had more than 10 videos posted during the relevant campaign period. In total, we collected YouTube channels for 17 politicians and 10 parties.

To identify YouTube channels of national news media, we began with the Media Bias/Fact Check list, which evaluates sources based on their editorial content and reporting practices~\cite{mediabias}. As this list was incomplete for French outlets, we supplemented it with sources recognized by the Conseil supérieur de l’audiovisuel (CSA) France’s former regulatory authority for audiovisual media ~\cite{CSA}. We included only channels that consistently published informational content, not solely entertainment, over a three-month period and had at least 10,000 subscribers. In total, we identified channels of 36 national news sources that met our requirements.


For local news channels, we used the list provided by the Alliance pour les Chiffres de la Presse et des Médias ~\cite{ACPM}, which tracks circulation and audience data for regional print and online media. The list included 62 regional outlets, of which we located YouTube channels for 43. We selected those with over 5,000 subscribers and at least some informational content published over a six-month period. In total, 15 regional channels met our criteria and were included for the next step of data collection. 

We identified the political orientations of politicians and their affiliated parties using Légifrance, the official French government website for legal information~\cite{Legifarnce}. Determining the political orientation of news channels was challenging due to limited official information and the evolving nature of media bias. To support our classifications, we relied on references from Media Bias/Fact Check and the European media monitoring platform Eurotopics~\cite{eurotop}.

 

After finalizing the selection of YouTube channels, we utilized the YouTube API to gather detailed channel metadata, including the number of videos, subscribers, and total likes, to assess the scale and engagement of each channel~\cite{youtube-api}. Using the same API, we retrieved video IDs for all videos published by these channels within the timeframe from March 1, 2023, to July 15, 2024 (to monitor time periods with and without elections). Additionally, we collected the metadata of each video: title, description, view count, like count, comment count, and publication date. \change{Metadata was collected in September, about two months after the last video in the dataset was published.}

Next, using the video IDs and the YouTube Transcript API,
we retrieved video transcripts~\cite{transcript-api}. Manually created transcripts were prioritized over automatically generated ones, and they were downloaded when available. In cases they were unavailable, auto-generated transcripts were collected. For videos without any available transcripts, no further attempts were made to obtain them. 
Since the data was publicly available and did not involve private individuals or sensitive information, individual consent was not required under current research ethics guidelines.
From party and politician channels (for simplicity, we refer to them as political channels throughout the paper), we collected 4,678 transcripts out of 5,435 videos (\textbf{Political dataset}). For national news media channels, we obtained 100,582 transcripts from a total of 150,843 videos (\textbf{News dataset}), and for regional news media, we collected 4,657 transcripts from 8,873 videos (\textbf{Local dataset}). The considered channels and their statistics are shown in Appendix Tables~\ref{tab:len_counts_grouped}, \ref{tab:len_counts_grouped_polit} and \ref{tab:local_subs}. Some transcripts were unavailable, either because the videos contained no spoken language or because the channels disabled auto-captioning and did not provide their own. Even though some transcripts are missing, the dataset still includes many videos from each main political group, making meaningful comparisons possible.

\section{Methodology for Themes Extraction}\label{data_class}
\begin{figure}[t]
  \centering
  \includegraphics[width=0.47\textwidth]{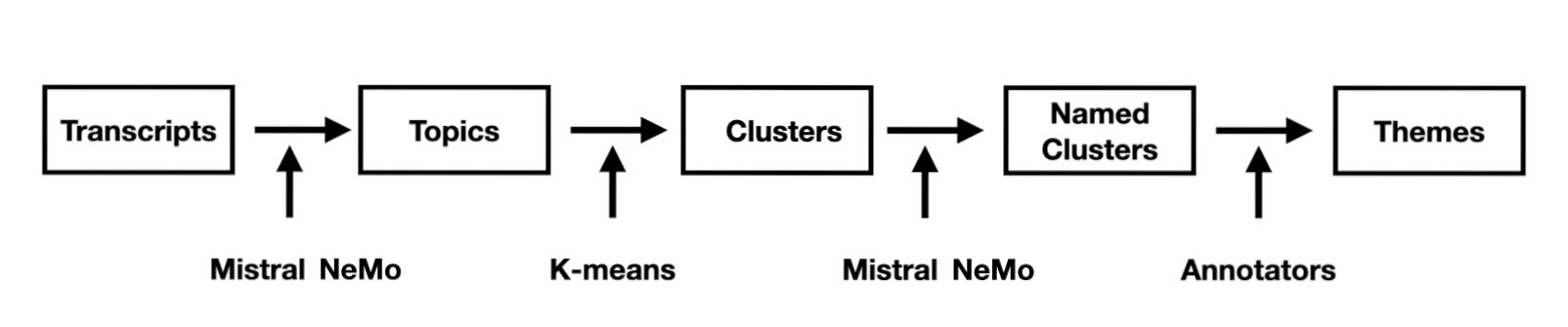}
  \caption{Themes extraction pipeline.}
  \label{fig:sample-figure}
\end{figure}

This section outlines the theme extraction process (Figure \ref{fig:sample-figure}), including using a language model to identify topics, the clustering method applied, and the final thematization step grouping related clusters into broader themes.

\change{We started our theme extraction with a classic topic modeling approach using BERTopic~\cite{bertopic}. To improve granularity, we split transcripts into 500-word segments and ran the algorithm multiple times. However, the results were unsatisfactory: all topics related to French politics were collapsed into a single category, and many of the extracted topics were difficult to interpret.}

\change{In parallel, recent studies have shown that large language models (LLMs) can be effectively employed for topic extraction~\cite{mu2024largelanguagemodelsoffer, doi-etal-2024-topic, Invernici_2025, CELIKTEN2025113219}.}
To perform topic labeling on our transcripts, we selected a suitable large language model (LLM) based on four main criteria. The model needed strong performance on labeling tasks to ensure accuracy, and training on French-language data to match our transcripts. We prioritized open-source models for flexibility and transparency, and considered cost-efficiency, given the computational demands of large-scale processing. Considering all these aspects, we selected NeMo~\cite{nemo}, an LLM developed by Mistral.


We chose not to rely on a predefined list of topics for data labeling, as this approach enables us to analyze and label data without requiring prior knowledge of its content. By avoiding predefined topics, the labeling process can reveal unexpected themes or insights that might otherwise be overlooked.
We experimented with several prompts to identify the most effective one (see Appendix Table~\ref{tab:prompt_exmp}).
Without a specified number of topics, the model often produced too many, including redundant or less relevant topics.
We adopted the following approach: transcripts under 500 words were assigned one topic; longer texts received one additional topic per 500 words (up to five for 2,000–12,000 words). Transcripts over 12,000 words were split into 1,000-word segments, with two topics per segment to address input size limits. This ensured more balanced and context-sensitive topic extraction. 
\change{We also tested two setups: using prompts in French and using prompts in English while requesting topics in French. In more than 80\% of cases, the resulting topics were either identical (after translation into English) or semantically very close (cosine similarity above 0.7). However, in about 6\% of cases, the topics produced in French, or requested in French, were difficult to interpret. As a result, we decided to rely on English prompts requesting topics in English.} An example of a prompt for texts under 500 words is (\change{the full prompt is provided in the appendix}):

\textit{``You are given a transcript of the video. Detect one main topic from the given transcript. All topics should be a maximum of three words and in English.''}

\change{The common issues appeared in videos with very short or text-poor transcripts, where the limited linguistic information made reliable topic extraction difficult. This problem was especially noticeable in Shorts or visually oriented content. To illustrate these cases, we provide translated transcripts below (the original language was French).
For example, here is a transcript of far-right politician Bardella's video:
\textit{“Yeah, but it's... I'm really embarrassed to be saying my thing in front of an empty room. We're here, we're here. Okay Jordan, okay Jordan, will you marry me? We have seven days left, seven days left.”}
was labeled by the model as “public speaking anxiety.” While the transcript mentions a speech, the actual content was intended humorously, leading the model to produce an overly literal interpretation.
A second case comes from a cooking video by Le Point with the transcript:
\textit{“Today's little tip is that instead of throwing away the peelings, we're going to make a soup. To make it even better, we're going to add four asparagus spears. We wash everything, blanch it quickly in water, mix it with a little cooking water, a drizzle of olive oil, and enjoy.”}
which was classified as “waste reduction.” Although technically related, the main focus of the video was a recipe rather than environmental sustainability.
These cases illustrate the model’s tendency to overgeneralize short or informal transcripts into more serious thematic categories.}

After obtaining topics, we used Mistral embeddings to represent the topics in a high-dimensional space and applied k-means clustering to group similar topics~\cite{ms_embedding}. 
\change{To determine the number of clusters, we tested the elbow method and silhouette analysis. Although these methods produced reasonable results, they tended to merge distinct topics that we deemed should be kept separate  (such as \textit{Macron's Governance} and \textit{Zemmour}). Since the dataset can be meaningfully partitioned at different levels of granularity, we opted for a larger number of clusters to preserve fine distinctions and allow for flexibility for later aggregation into broader themes.}




We manually evaluated the 30 largest and 30 smallest clusters from each dataset to validate the clustering quality. Our assessment focused on two main criteria: internal thematic coherence and sufficient topic density. Clusters were expected to contain topically related items that reflected a clear and consistent theme while being substantial enough to avoid over-fragmentation. Specifically, we aimed to minimize clusters that were either incoherent, containing loosely related topics, or overly granular, consisting of only one or two topics. If many large clusters appeared incoherent, we increased the number of clusters to improve specificity. Conversely, if too many small clusters were too narrow, we reduced the number to encourage more meaningful groupings. This iterative approach allowed us to strike a balance between coherence and interpretability.


After obtaining the clusters, we applied Mistral's NeMo model to assign representative names to them. The model took all topics within each cluster as input and returned a name that reflected the dominant theme or majority focus of the cluster. We manually reviewed a random sample of 50 clusters and their assigned names to ensure that the labels accurately reflected each cluster's underlying topics and thematic focus. All reviewed cluster names were found to be appropriate, confirming the reliability of the naming process. 
Following this, we manually merged closely related clusters and referred to the resulting groups as themes. 
We avoided purely algorithmic approaches because they often produced semantically incoherent groupings, merging distinct topics based only on distributional similarity. Manual merging allows us to apply semantic understanding that current clustering algorithms cannot reliably capture.
This process was conducted collaboratively by two researchers, with each decision discussed and refined until full agreement was reached. Themes vary in their level of granularity. 
Some themes are broader in scope, such as \textit{Crimes \& Law}, which includes general clusters like \textit{Justice \& Law Enforcement} and \textit{Justice \& Law}, as well as clusters that do not clearly belong to a separate theme, such as \textit{Monique Olivier's Trial}. In contrast, more specific themes, like \textit{Sexual Misconduct \& Abuse}, focus on narrowly defined issues within a particular area. 

Decisions about merging were based on two main criteria: the popularity of clusters (i.e., whether multiple clusters addressed similar issues) and the specificity of the topics they included (i.e., whether they dealt with general or more focused subtopics). 
The finalized themes formed the basis of our video analysis. However, our dataset retained all intermediate results, allowing us to trace the hierarchy from broader themes to  more fine-grained topics.

To evaluate the quality of our final themes, we manually reviewed 100 videos from the news dataset and 50 videos from the political and local datasets. Each video selected for review was no longer than 1000 words. For each video, we asked two questions: \textit{(1) Are the assigned themes accurate given the video’s content?} and \textit{(2) Do the themes cover all topics discussed in the video?}
\change{These questions were answered in a binary fashion (Yes/No), with ``Yes'' meaning that the assigned themes correctly represented the main content (Question 1) or sufficiently covered all major themes (Question 2).  In cases of partial correctness (e.g., one theme correct, another completely incorrect), the judgment was recorded as ``No''. Two researchers performed the evaluation independently. They initially agreed on more than 90\% of the transcripts, and any uncertainties were discussed until agreement was reached.} The results indicated that 90\% of the videos had accurately assigned themes, while 74\% included all relevant topics discussed beyond just the main ones. 

\change{We assessed theme coherence using cosine similarity between topic embeddings and their theme medoids, and distinctiveness through cross-theme comparisons. The intra-theme analysis showed strong cohesion (average similarity = 0.75), while inter-theme similarity was lower (0.66). The highest inter-theme similarity appeared between \textit{Israel–Hamas War} and \textit{Middle East Conflicts} (0.75) which are closely related but address different aspects of international conflicts.
Thus, we treat them as distinct themes. In contrast, themes such as \textit{Celebrity \& Pop Culture} (0.67) and \textit{Sport} (0.68) showed the lowest intra-theme cohesion, as they encompass a wide range of individual names and figures.}

\section{Results}\label{results}


\subsection{Analysis of Theme Distribution}
\paragraph{News dataset.} First, we examine the most prominent themes within the News dataset during the whole period of data collection (Table~\ref{tab:news_ds_frq_themes}). 
Center-aligned news channels frequently focus on conflicts and geopolitical issues, covering both general topics and specific events such as the \textit{Russia-Ukraine War} and the \textit{Israel-Hamas War}. In contrast, left-leaning news channels primarily emphasize themes related to the \textit{News Media \& Freedom of Expression}, and \textit{Protests}. Right-wing news channels tend to emphasize themes related to \textit{Immigration}, \textit{Healthcare}, and \textit{Sports}.
These results suggest that news channels' thematic focus aligns with their political orientation: center-aligned channels emphasize international conflicts, left-leaning channels highlight civil liberties and social movements, and right-leaning channels focus on socio-political issues (\textit{Immigration} and \textit{Healthcare}).



\begin{table}[!htbp]
\small
\centering
\begin{tabular}{|l|l|r|r|}

\hline
\textbf{Political}& \multirow{2}{*}{\textbf{Theme}}& \multirow{2}{*}{\textbf{Occ.}}& \multirow{2}{*}{\textbf{\%} }\\
\textbf{orient.}& & & \\
\hline
\multicolumn{4}{|c|}{\textbf{Entire period}}\\
\hline
\multirow{5}{*}{Left} & Politics \& Polit. Figures & 718 & 29.64 \\
& News Media \& Freedom of Expr. & 472 & 19.49 \\
& Crimes \& Law & 462 & 19.08 \\
& Protests & 311 & 12.84 \\
& Israel–Hamas War & 308 & 12.72 \\
\hline
\multirow{5}{*}{Center} & Politics \& Polit. Figures & 7084 & 15.56 \\
& Crimes \& Law & 4711 & 10.35 \\
& Israel–Hamas War & 4056 & 8.91 \\
& Conflicts \& Geopolitics & 3970 & 8.72 \\
& Russia-Ukraine War & 3921 & 8.61 \\
\hline
\multirow{5}{*}{Right} & Politics \& Polit. Figures & 7865 & 16.65 \\
& Crimes \& Law & 7285 & 15.42 \\
& Health \& Healthcare & 4259 & 9.02 \\
& Sport & 3472 & 7.35 \\
& Immigration \& Refugees & 2716 & 5.75 \\

\hline
\multicolumn{4}{|c|}{\textbf{Pre-election period}} \\
\hline
\multirow{5}{*}{Left} & Politics \& Polit. Figures & 515 & 31.1 \\
& Crimes \& Law & 358 & 21.6 \\
& News Media & 298 & 17.9 \\
&Left-Wing Politics	&258 &15.9\\
& Protests & 233 & 14.1 \\
\hline
\multirow{5}{*}{Center} & Politics \& Polit. Figures & 4504 & 14.1 \\
& Crimes \& Law & 3394 & 10.6 \\
& Russia-Ukraine War & 2774 & 8.70\\
& Israel–Hamas War & 2699 & 8.5 \\
& Conflicts \& Geopolitics & 2647 & 8.3 \\
\hline
\multirow{5}{*}{Right} & Crimes \& Law & 4937 & 15.6 \\
& Politics \& Polit. Figures & 4773 & 15.1 \\
& Health \& Healthcare & 2831 & 8.1 \\
& Sport & 2435 & 7.7\\
& Immigration \& Refugees & 2043 & 6.5 \\
\hline
\multicolumn{4}{|c|}{\textbf{European election period}} \\
\hline
\multirow{5}{*}{Left}& Israel–Hamas War & 127 & 23.6 \\
& Politics \& Polit. Figures & 126 & 23.5 \\
& News Media& 118 & 22 \\
& Crimes \& Law & 86 & 16 \\
& Human Rights & 72 & 13.4 \\
\hline
\multirow{5}{*}{Center}  & Politics \& Polit. Figures & 1468 & 15 \\
 & Israel–Hamas War & 1185 & 12.1 \\
 & Crimes \& Law & 1054 & 10.77 \\
 & Conflicts \& Geopolitics & 1041 & 10.6 \\
 & Russia-Ukraine War & 912 & 9.3 \\
\hline
\multirow{5}{*}{Right} & Crimes \& Law & 1835 & 16.7 \\
 & Politics \& Polit. Figures & 1477 & 13.4 \\
& Health \& Healthcare & 1117 & 10.2 \\
& Russia-Ukraine War & 787 & 7.2 \\
 & Sport & 674 & 6.1 \\
\hline
\multicolumn{4}{|c|}{\textbf{Legislative election period}} \\
\hline
\multirow{5}{*}{Left}
& Left Wing Politics & 65 & 28.6 \\
& Far-Right Politics & 113 & 49.8 \\
& Politics \& Polit. Figures & 77 & 33.9 \\
&Election&58&25.5\\
& News Media  & 56 & 24.7 \\
\hline
\multirow{5}{*}{Center}& Politics \& Polit. Figures & 1112 & 28.8 \\
& Election & 859 & 22.2 \\
& Far-Right Politics & 592 & 15.31 \\
& Conflicts \& Geopolitics & 282 & 7.3 \\
& Left Wing Politics & 276 & 7.1 \\
\hline
\multirow{5}{*}{Right} & Politics \& Polit. Figures & 1615 & 34.9 \\
 & Election & 1027 & 22.2 \\
 & Far-Right Politics & 640 & 13.8 \\
 & Crimes \& Law & 513 & 11.1 \\
 & Macron's Presidency & 496 & 10.7 \\
\hline
\end{tabular}
\caption{\change{\textbf{News dataset:} Most frequent themes by period.}}
\label{tab:news_ds_frq_themes}
\end{table}

Next, we analyze the most frequent themes across three different political time periods: (i) \textbf{pre-election period}, from March 1, 2023 to March 1, 2024, (ii) \textbf{European election period} from March 1, 2024 to June 8, 2024, and (iii) \textbf{legislative election period}, from June 8, 2024 to July 15, 2024.
The distribution of themes during the pre-election period closely resembles that of the overall dataset, as this period accounts for approximately 70\% of the total data collection time frame (Table~\ref{tab:news_ds_frq_themes}). However, during the European election period, notable shifts in theme coverage have emerged (Table~\ref{tab:news_ds_frq_themes}). For instance, the \textit{Israel-Hamas War} was the most frequent theme among left-leaning news channels and the second most frequent among center-aligned channels. Interestingly, this theme does not appear among the top five most frequent themes in right-leaning news channels. The \textit{Election} theme did not rank among the top five themes of any news channel group during the European election period. It ranked sixth for center-aligned channels, seventh for right-leaning channels, and eleventh for left-leaning channels. Right-leaning news channels paid relatively more attention to election-related topics compared to left-leaning channels. This focus on the election aligns with the stronger performance of right-wing parties in the 2024 European Parliament election in France.

During the legislative election period, the \textit{Election} theme ranked among the top five themes across all political orientations, alongside the \textit{Far-Right Politics} theme. The \textit{Left-Wing Politics} theme was prominent among center- and left-leaning news channels but did not feature in the top five for right-leaning channels. In addition, some earlier patterns persisted: \textit{Conflicts \& Geopolitics} remained among the top five themes for center-aligned channels, \textit{News Media \& Freedom of Expression} for left-leaning channels, and \textit{Crimes \& Law} for right-leaning channels (Table~\ref{tab:news_ds_frq_themes}).

\begin{table}[!htbp]
\centering
\small
\renewcommand{\arraystretch}{1.1}
\begin{tabular}{|l|l|r|r|}
\hline
\textbf{Political}& \multirow{2}{*}{\textbf{Theme}}& \multirow{2}{*}{\textbf{Occ.}}& \multirow{2}{*}{\textbf{\%} }\\
\textbf{orient.}& & & \\
\hline
\multirow{5}{*}{Left} & Politics \& Polit. Figures & 808 & 27.8 \\
&Economy \& Market Dynamics & 573 & 19.68	\\
& Election & 571 & 19.6 \\
& Climate Change \& Env. Problems & 482 & 16.6 \\
&Left-Wing Politics	&447&15.4\\
\hline
\multirow{5}{*}{Center} & Politics \& Polit. Figures & 159 & 26.2 \\
& Climate Change \& Env. Problems & 95 & 15.6 \\
&Economy and Market Dynamics&94&15.5\\	
& European Union & 80 & 13.7 \\
& Election & 79 & 13 \\
\hline
\multirow{5}{*}{Right} & Politics \& Polit. Figures & 122 & 50 \\
& European Union & 105 & 43 \\
& Election & 82 & 33.6 \\
& News Media \& Freedom of Express. & 68 & 27.9 \\
& Macron's Presidency & 53 & 21.7 \\
\hline
\multirow{5}{*}{Far-Right} & European Union & 345 & 30.3 \\
& Politics \& Polit. Figures & 290 & 25.4 \\
& Election & 262 & 23 \\
& Protests & 239 & 21 \\
& Macron's Presidency & 213 & 18.7 \\
\hline
\end{tabular}
\caption{\textbf{Political dataset:} Most frequent themes during the entire period.}
\label{tab:polit_dataset_all}
\end{table}

\paragraph{Political dataset.} In the Political dataset, the diversity of themes is more limited compared to the News dataset (Table~\ref{tab:polit_dataset_all}). Themes such as \textit{Election} and \textit{Politics \& Political Figures} consistently appear in the top five across all political groups. The \textit{European Union} theme also ranks among the top five for all groups, with the exception of left-wing politicians. \textit{Climate Change \& Environmental Problems} is the second most frequent theme for center-aligned political channels, and the fourth most frequent for left-wing political actors.
However, it does not appear among the top five most frequent themes for right- and far-right politicians. For both right- and far-right political figures, the \textit{Macron's Presidency} theme ranks as the fifth most frequent during the data collection period. Interestingly, it does not appear among the top five themes for center-aligned politicians. The theme \textit{News Media \& Freedom of Expression} is among the most frequently discussed by right-wing politicians.

\begin{table}[!htbp]
\centering
\small
\begin{tabular}{|l|l|r|r|}
\hline
\textbf{Political}& \multirow{2}{*}{\textbf{Theme}}& \multirow{2}{*}{\textbf{Occ.}}& \multirow{2}{*}{\textbf{\%} }\\
\textbf{orient.}& & & \\
\hline
\multicolumn{4}{|c|}{\textbf{Pre-election period}} \\
\hline
\multirow{5}{*}{Left}
& Politics \& Polit. Figures & 489 & 30 \\
&Economy \& Market Dynamic &332&20.0\\
& Climate Change \& Env. Problems & 296 & 18 \\
& Protests & 273 & 16.5 \\
& Pension Reform & 266 & 16.1 \\
\hline
\multirow{5}{*}{Center}
& Politics \& Polit. Figures & 117 & 24.6 \\
& Climate Change \& Env. Problems & 83 & 17.5 \\
&Economy \& Market Dynamics &71&14.95\\
& Social \& Economic Inequality & 61 & 12.8 \\
& Pension Reform & 55 & 11.6 \\
\hline
\multirow{5}{*}{Right}
& Politics \& Polit. Figures & 51 & 53.1 \\
& European Union & 35 & 36.5 \\
& News Media & 30 & 31.2 \\
& Immigration \& Refugees & 23 & 24 \\
& Macron's Presidency & 23 & 24 \\
\hline
\multirow{5}{*}{Far-Right}
& European Union & 201 & 33.8 \\
& Protests & 186 & 31.3 \\
& Politics \& Polit. Figures & 150 & 25.2 \\
 & News Media & 127 & 21.3 \\
 & Macron's Presidency & 118 & 19.8 \\
\hline
\multicolumn{4}{|c|}{\textbf{European election period}} \\
\hline
\multirow{5}{*}{Left} & Election & 262 & 28.5 \\
& Israel–Hamas War & 209 & 22.8 \\
&Economy \& Market Dynamics &197 & 21.46\\
& European Union & 187 & 20.4 \\
& Politics \& Polit. Figures & 186 & 20.3 \\

\hline
\multirow{5}{*}{Center} & Election & 39 & 35.4 \\
& European Union & 31 & 28.2 \\
& Politics \& Polit. Figures & 27 & 24.5 \\
& Economy \& Market Dynamics & 19 & 17.3 \\
& Russia-Ukraine War & 13 & 11.8 \\
\hline
\multirow{5}{*}{Right} & European Union & 43 & 59.7 \\
& Politics \& Polit. Figures & 37 & 51.4 \\
& Election & 26 & 36.1 \\
& News Media & 20 & 27.8 \\
& France's Intl. Relations & 16 & 22.2 \\
\hline
\multirow{5}{*}{Far-Right} & Election & 130 & 31.4 \\
& European Union & 126 & 30.4 \\
& Politics \& Polit. Figures & 98 & 23.7 \\
& Immigration \& Refugees & 72 & 17.4 \\
& Macron's Presidency & 71 & 16.7 \\
\hline
\multicolumn{4}{|c|}{\textbf{Legislative election period}} \\
\hline
\textbf{Political}& \multirow{2}{*}{\textbf{Theme}}& \multirow{2}{*}{\textbf{Occ.}}& \multirow{2}{*}{\textbf{\%} }\\
\textbf{orient.}& & & \\
\hline
\multirow{5}{*}{Left} 
&Election Politics&138&41.1\\
& Politics \& Polit. Figures & 133 & 39.6 \\
&Left-Wing Politics&126&37.5\\
& Far-Right Politics & 75 & 24.3 \\
& Social \& Economic Inequality & 66 & 21.4 \\
\hline
\multirow{5}{*}{Center} & Politics \& Polit. Figures & 15 & 65.2 \\
& Election & 12 & 52.2 \\
& Far-Right Politics & 10 & 43.5 \\
& Social \& Economic Inequality & 5 & 21.7 \\
& Macron's Presidency  & 4 & 17.4 \\
\hline
\multirow{5}{*}{Right} & Election & 37 & 48.7 \\
& Politics \& Polit. Figures & 34 & 44.7 \\
& European Union & 27 & 35.5 \\
& News Media  & 18 & 23.7 \\
& Macron's Presidency & 16 & 21 \\
\hline
\multirow{5}{*}{Far-Right} & Election & 62 & 47.3 \\
& Politics \& Polit. Figures & 42 & 32.1 \\
& Far-Right Politics & 26 & 19.9 \\
& Macron's Presidency & 24 & 18.3 \\
& European Union & 18 & 13.7 \\
\hline

\end{tabular}
\caption{\change{\textbf{Political dataset:} Most frequent themes by period.}}
\label{tab:plt_common_themes_orientation}
\end{table}




To better understand how thematic focus shifted over time, we examined specific periods in greater detail (Table \ref{tab:plt_common_themes_orientation}). During the European election period, the \textit{Election} and the \textit{European Union} themes were the most prominent across all political groups in the Political dataset. In contrast, these themes did not rank among the top five in the News dataset during the same time frame. Center-aligned political actors focus more on \textit{Economy \& Market Dynamics} and \textit{Russia-Ukraine War}, showing more interest in global and economic issues. Left-leaning figures pay more attention to \textit{Israel–Hamas War}, reflecting concern for humanitarian topics. Right-leaning actors highlight \textit{France’s International Relations} and \textit{News Media \& Freedom of Expression}, indicating a focus on national image and media coverage. The far right, on the other hand, emphasized \textit{Immigration \& Refugees} and \textit{Macron’s Presidency}, reflecting their main political topic and oppositional stance towards the French President.

During the legislative election period, a notable shift is the rise of \textit{Far-Right Politics} as one of the most frequently discussed themes among center, left, and far-right political figures (Table  \ref{tab:plt_common_themes_orientation}).
This is obviously due to the success of the Rassemblement National at the European elections, which in turn led Macron to call new legislative elections in France. 
In contrast, \textit{Left-Wing Politics} was primarily discussed by left-leaning actors, while \textit{Macron’s Presidency} emerged as a major theme across all political groups except the left. We also observe that \textit{Social \& Economic Inequality} is an important topic for the left and center but not for the right and far right whereas the \textit{European Union} is still extensively discussed in these last two groups.
Throughout the analysis of the political dataset, the \textit{News Media \& Freedom of Expression} theme consistently appears among the top five most frequent themes for right-leaning channels.
Overall, the data points to a shifting political landscape in which topics related to far-right politics moved into the mainstream of political debate.

Political channels were significantly more active during the European and legislative election periods. Although these elections spanned only 4.5 months, representing 27\% of the total data collection time frame, they accounted for 42\% of all videos published by political channels during the entire period. This pattern does not extend to news channels, which published 30\% of their videos during the European and legislative election periods, only slightly higher than the 27\% these periods represent in the overall data collection time frame.




\paragraph{Local dataset.} We also included an analysis of local news sources, as prior studies indicate that people tend to place greater trust in local news providers compared to national or official outlets~\cite{local}. Given that the dataset includes only 4,000 videos from local sources, we chose not to divide them by time periods or political orientation.
\begin{table}[h!]
\centering
\small
\begin{tabular}{|l|r|r|}
\hline
\textbf{Theme} & \textbf{Occ.} & \textbf{\%} \\
\hline
Sport & 1100 & 23.6 \\
Economy \& Market Dynamics & 456 & 9.8 \\
Crimes \& Law & 388 & 8.3 \\
Local News & 320 & 6.9 \\
Festivals \& Celebrations & 313 & 6.7 \\
\hline
\end{tabular}
\caption{\textbf{Local dataset:} Most frequent themes.}
\label{tab:top_themes_summary}
\end{table}

The thematic distribution in the local dataset highlights the strong community focus of local news content and less emphasis on political content (Table \ref{tab:top_themes_summary}). \textit{Sport} is the most frequent theme, accounting for 23.6\% of all occurrences. This is followed by \textit{Economy \& Market Dynamics} and \textit{Crimes \& Law}, which reflect practical concerns that directly impact local communities. Themes such as \textit{Local News} and \textit{Festivals \& Celebrations} further underscore the focus on everyday life and community engagement. Overall, the local dataset reflects a strong emphasis on community-centered content, with a focus on sports, everyday concerns, and local events.

\change{\subsection{Topics Analysis of a Notable Theme}}

\change{Since the theme \textit{News Media \& Freedom of Expression} ranks in the top five for both right politicians and left-leaning news media channels, we analyzed what topics emerged within this broader theme and how each group emphasized different aspects.}

\change{The political dataset reveals that the \textit{Media Bias} topic appears across all four political categories, but each group emphasizes different aspects of the topic. Left-leaning politicians exhibit the strongest tendency to discuss media-related issues, particularly focusing on \textit{Journalist's Behavior}, \textit{Media Independence}, and \textit{Press Freedom}. Center politicians engage less frequently with media topics overall, but when they do, their discussions center on media bias and misinformation, with topics like \textit{Misinformation}, \textit{Media Freedom} and \textit{Information Manipulation}.
Right-leaning channels demonstrate significant engagement with media bias themes, especially \textit{Media Bias in Elections}, reflecting their focus on media's political influence. Far-right politicians prioritize freedom of speech concerns, addressing topics like \textit{Censorship in the media} and \textit{Petition Against Censorship}.

In the news dataset, \textit{Media Bias} and \textit{Freedom of Speech} appear across all categories of channels, though their emphases differ. Left-leaning outlets primarily focus on press freedom and the role of journalism, often highlighting topics such as \textit{Investigative Journalism} and \textit{Media and Journalism}. Center-leaning outlets resemble the left in their strong emphasis on press freedom, but they also put an emphasis on media regulation, particularly in discussions of \textit{Media Regulation} and \textit{Public Broadcasting Reform}. Right-leaning outlets, by contrast, similarly to the discourse of far-right politicians focus heavily on freedom of speech. While this topic appears across all channels, in right-leaning news media it represents 9\% of discussions, highlighting its particular prominence.}

\subsection{Analysis of Engagement Metrics}
To further understand audience interest in the different themes covered by news and political channels, we analyze the key engagement metrics, specifically the like-to-view and comment-to-view ratios. These measures provide insight into how actively viewers responded to content beyond simply watching it.
To calculate the highest average comment-to-view and like-to-view ratios for each political group of news channels, we considered only themes that appeared at least ten times within that group.

\begin{table}[h!]
\centering
\small
\begin{tabular}{|l|l|r|r|}
\hline
\textbf{Political}& \multirow{2}{*} {\textbf{Theme} }&\textbf{Comm/}& \multirow{2}{*} {\textbf{Occ.} }\\
{\textbf{Orient.}}&&\textbf{View}&\\

\hline
\multirow{3}{*}{Left} & Religion \& Belief & 0.0097 & 23 \\
& Left-Wing Politics & 0.0096 & 137 \\
& Russian Politics & 0.0091 & 18 \\
\hline
\multirow{3}{*}{Center} & Antisemitism & 0.0094 & 413 \\
& Left-Wing Politics & 0.0080 & 541 \\
&Far-Right Politics	&0.0072	&1212 \\
\hline
\multirow{3}{*}{Right} & Antisemitism & 0.0129 & 983 \\
& Islam \& French Society & 0.0115 & 461 \\
& Left-Wing Politics & 0.0098 & 909 \\
\hline
\end{tabular}
\caption{\textbf{News dataset:} highest comment-to-view ratios. Occ. refers to the number of videos including the theme.}
\label{tab:news_comment_view_ratio}
\end{table}

\paragraph{News dataset.} For the News dataset, themes with the highest comment-to-view ratios tend to be more polarizing, with \textit{Antisemitism} ranking highest for both center- and right-leaning political groups (Table~\ref{tab:news_comment_view_ratio}). The \textit{Left-Wing Politics} theme was widely discussed across all political groups.
Additionally, right-leaning channels demonstrated the highest overall comment engagement for their top themes.
\change{However, on average, left-leaning news outlets exhibit the highest comment-to-view ratio (0.007), compared to 0.005 for right-leaning channels and 0.004 for center-leaning channels.}
 
\begin{table}[h!]
\centering
\small
\renewcommand{\arraystretch}{1.1}
\begin{tabular}{|l|l|r|r|}
\hline
\textbf{Political}& \multirow{2}{*} {\textbf{Theme} }&\textbf{Like/}& \multirow{2}{*} {\textbf{Occ.} }\\
\textbf{Orient.}&&\textbf{View}&\\
\hline
\multirow{3}{*}{Left} & Nature \& Wildlife	&0.0579	&12\\
&Olympic Games & 0.0462 & 32 \\
& Food Industry \& Regulations & 0.0460 & 35 \\
\hline
\multirow{3}{*}{Center} & Video Games & 0.0166 & 82 \\
& Animal Welfare \& Rights & 0.0151 & 124 \\
& Streaming \& Content Creation & 0.0150 & 73 \\
\hline
\multirow{3}{*}{Right} & Animal Welfare \& Rights&0.0189&297\\
&Agriculture&0.0184&1469\\
&European Union&0.0184&1209\\
\hline
\end{tabular}
\caption{\textbf{News dataset:} highest like-to-view ratios. 
Occ. refers to the number of videos including the theme.}
\label{tab:news_like_view_ratio}
\end{table}

\change{In terms of like-to-view ratios, left-leaning news outlets again have the highest level (0.04), followed by right-leaning channels (0.012) and center channels (0.01), suggesting higher overall audience approval.}
Contrary to the comment-to-view ratios, the like-to-view ratios tend to be higher for non-controversial themes (Table~\ref{tab:news_like_view_ratio}). For example, among center- and left-leaning news channel groups, the top themes are unrelated to politics, with \textit{Video Games} leading for the center group and \textit{Nature} for the left-leaning group. The situation is different for right-leaning channels: politically charged themes, the \textit{European Union} and \textit{Agriculture} (primarily related to farmer protests) rank among the themes with the highest like-to-view ratio. 

\begin{table}[h!]
\centering
\small
\begin{tabular}{|l|l|r|r|}
\hline
\textbf{Political}& \multirow{2}{*} {\textbf{Theme} }&\textbf{Comm/}& \multirow{2}{*} {\textbf{Occ.} }\\
{\textbf{Orient.}}&&\textbf{View}&\\
\hline
\multirow{3}{*}{Left} & Military & 0.0287 & 41 \\
& Russia-Ukraine War & 0.0209 & 152 \\
& European Union & 0.0196 & 307 \\
\hline
\multirow{3}{*}{Center} 
& Agriculture & 0.0118 & 37 \\
& Russia-Ukraine War & 0.0116 & 38 \\
&Economy &0.0084&67\\
\hline
\multirow{3}{*}{Right} & Agriculture & 0.0133 & 12 \\
& Election & 0.0129 & 82 \\
& Far-Right Politics & 0.0127 & 16 \\
\hline
\multirow{3}{*}{Far-Right} & Terrorism \& Security & 0.0175 & 10 \\
& Olympic Games & 0.0151 & 10 \\
& Far-Right Politics & 0.0139 & 71 \\
\hline
\end{tabular}
\caption{\textbf{Political dataset:} highest comment-to-view ratios. Occ. refers to the number of videos including the theme.}
\label{tab:comment_view_ratio_orientation}
\end{table}

\paragraph{Political dataset.}
\change{In the political dataset, left-leaning channels have the highest average comment-to-view ratio (0.014), similarly to the news dataset. Among politicians, far-right and right-leaning figures have average ratios of 0.012 and 0.01, respectively, while center-leaning politicians record the lowest value at 0.005.}
For left-wing political actors, the themes with the highest comment-to-view ratios are  \textit{Military}, \textit{Russia-Ukraine War}, and \textit{European Union}, suggesting strong engagement with international and institutional issues (Table \ref{tab:comment_view_ratio_orientation}). Content from centrist political figures shows the lowest comment-to-view ratios, even when addressing polarizing topics, suggesting a more passive or less reactive audience. Right-leaning content, including themes like \textit{Agriculture} and \textit{Election}, demonstrates moderate but consistent engagement, reflecting a steady level of audience interaction. Meanwhile, far-right political content sees elevated comment activity on themes such as \textit{Terrorism \& Security} and \textit{Far-Right Politics}, pointing to heightened responsiveness to polarizing narratives within this audience.

\change{The highest average like-to-view ratio is observed among far-right politicians (0.09), followed by right-leaning (0.07) and left-leaning politicians (0.06), whereas center-leaning politicians show the lowest ratio (0.02).} We observe a greater presence of non-political content among top-ranked themes in terms of like-to-view ratio (Table \ref{tab:like_view_ratio_orientation}). For instance, \textit{Olympic Games} are within the top three themes for far-right and left-wing political channels, while \textit{French History \& Culture} ranks second for centrist channels. Nonetheless, politically relevant themes still appear, for example, \textit{Election} and \textit{Human Rights} for right-wing channels, \textit{Israel-Hamas War} for left-wing channels, and \textit{French International Relations} for centrist ones. This contrasts with the news dataset, where most highest like-to-view ratios are primarily associated with consensual or apolitical themes. Generally speaking, we observe that far-right politicians have the highest like-to-view ratios among political channels, unlike the news dataset where left-leaning channels have the highest audience approval. In both cases, centrist channels generate the lowest ratios, indicating a more passive or less mobilized audience base.

\begin{table}[h!]
\centering
\small
\begin{tabular}{|l|l|r|r|}
\hline
\textbf{Political}& \multirow{2}{*} {\textbf{Theme} }&\textbf{Like/}& \multirow{2}{*} {\textbf{Occ.} }\\
\textbf{Orient.}&&\textbf{View}&\\
\hline
\multirow{3}{*}{Left} 
& Food Industry  & 0.0643 & 29 \\
&Olympic Games&0.0625&12\\
& Israel–Hamas War & 0.0625 & 425 \\
\hline
\multirow{3}{*}{Center} & France's Int. Relations & 0.0259 & 18 \\
& French History \& Culture & 0.0253 & 13 \\
& Climate Change \& Env. Prob. & 0.0232 & 95 \\
\hline
\multirow{3}{*}{Right} & News Media & 0.0777 & 68 \\
& Election & 0.0761 & 82 \\
& Human Rights & 0.0738 & 18 \\
\hline
\multirow{3}{*}{Far-Right} & Health \& Healthcare & 0.1366 & 139 \\
& Olympic Games & 0.1341 & 10 \\
& Tech \& Innovation & 0.1330 & 42 \\
\hline
\end{tabular}
\caption{\textbf{Political dataset:} highest like-to-view ratios, where Occ. refers to the number of videos including the theme.}
\label{tab:like_view_ratio_orientation}
\end{table}

\paragraph{Local dataset.}
\change{In the local dataset, the average comment-to-view ratio is 0.002, while the average like-to-view ratio is 0.01.}
The themes with the highest comment-to-view ratios in the local dataset (\textit{Far-Right Politics}, \textit{European Union}, and \textit{Immigration}) reflect topics that are politically sensitive and often polarizing (Table \ref{tab:local_comment_view_ratio}). For the like-to-view ratio, we can see that non-political or issue-based themes such as \textit{Video Games} and \textit{AI \& Technology} elicit the strongest audience approval (Table \ref{tab:local_like_view_ratio}). The engagement metrics of local news sources more closely resemble those of national news channels than those of political figures, despite substantial differences in the composition of the datasets. Interestingly, both\textit{ Video Games} and \textit{Agriculture} also appear in the highest like-to-view themes in center and right (national) news channels, respectively.

\begin{table}[h!]
\centering
\small
\begin{tabular}{|l|c|r|}
\hline
\multirow{2}{*}{\textbf{Theme}} & \textbf{Comm/} & \multirow{2}{*}{\textbf{Occ.}} \\
                                & \textbf{View} &                                \\
\hline
Far-Right Politics & 0.010062 & 66 \\
European Union     & 0.007159 & 77 \\
Immigration        & 0.006820 & 73 \\
\hline
\end{tabular}
\caption{\textbf{Local dataset:} highest comment-to-view ratios. Occ. refers to the number of videos including the theme.}
\label{tab:local_comment_view_ratio}
\end{table}

\begin{table}[h!]
\centering
\small
\begin{tabular}{|l|c|r|}
\hline
\multirow{2}{*}{\textbf{Theme}} & \textbf{Like/} & \multirow{2}{*}{\textbf{Occ.}} \\
                                & \textbf{View}  &                                \\
\hline
Video Games & 0.039611 & 17 \\
Agriculture & 0.019132 & 137 \\
AI \& Technology & 0.018037 & 45 \\
\hline
\end{tabular}
\caption{\textbf{Local dataset:} highest like-to-view ratios. Occ. refers to the number of videos including the theme.}
\label{tab:local_like_view_ratio}
\end{table}

\subsection{Thematic Landscape of YouTube Channels}

To explore whether YouTube channels of news outlets and political figures tended to align along editorial lines, we visualized their thematic representations. We selected channels that have twenty or more videos in our dataset. For each channel, we constructed a theme vector by counting how often each theme appeared across its videos, counting all unique themes per video. These raw counts were then normalized to produce a probability distribution over themes for each channel. We then applied the dimensionality reduction technique t-SNE to project the theme vectors into a two-dimensional space for visualization, as it is well-suited for preserving local structure and revealing clustering patterns in high-dimensional data~\cite{tsne}.
The appendix provides details on the validation of the visualization results.

\begin{figure}[t]
  \centering
  \includegraphics[width=0.47\textwidth]{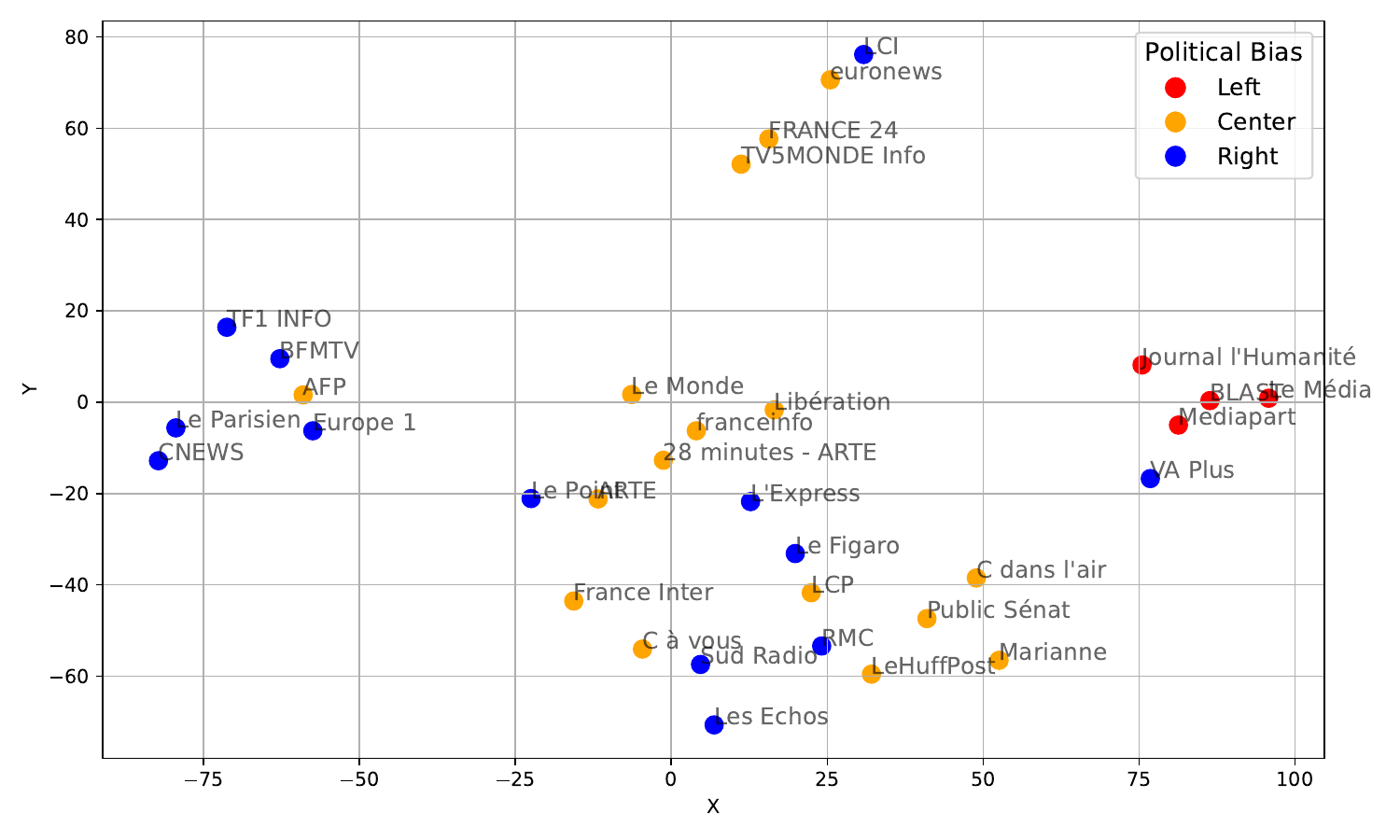}
  \caption{\textbf{News Dataset:} the t-SNE visualization of channel thematic coverage similarity.}
  \label{fig:news_vis}
\end{figure}

\paragraph{News dataset.}In the news dataset, the t-SNE projection reveals clear patterns in thematic profiles by political orientation (Figure \ref{fig:news_vis}). Left-leaning channels form a tight cluster, reflecting strong thematic coherence. VA Plus surprisingly appears close to this group, sharing themes such as \textit{News Media \& Freedom of Expression} and \textit{Protests} that are common in left-wing outlets.

In contrast, right and center-aligned channels are more widely dispersed, suggesting greater thematic diversity. A cluster of LCI, Euronews, TV5MONDE, and France 24 (mostly consisting of center channels) shows similarities tied to their international focus (with \textit{Russia-Ukraine War}, \textit{Israel-Hamas War}, and \textit{Conflicts \& Geopolitics} being the top ten most frequent ones). Another group, AFP, CNEWS, TF1 INFO, BFMTV, and Le Parisien, clusters around coverage of the \textit{Health \& Healthcare} theme.

The central cluster in the t-SNE plot, which includes outlets like Le Monde, franceinfo, Libération, ARTE, and L’Express, combines a range of well-established, widely read French news media.

Overall, the t-SNE visualizations reveal clear thematic polarization aligned with political orientation. Left-leaning actors and outlets cluster tightly, while center and right-wing entities are more dispersed, reflecting broader thematic diversity. Some figures, like VA Plus,  appear in unexpected positions due to specific thematic focus. 


\begin{figure}[t]
  \centering
  \includegraphics[width=0.47\textwidth]{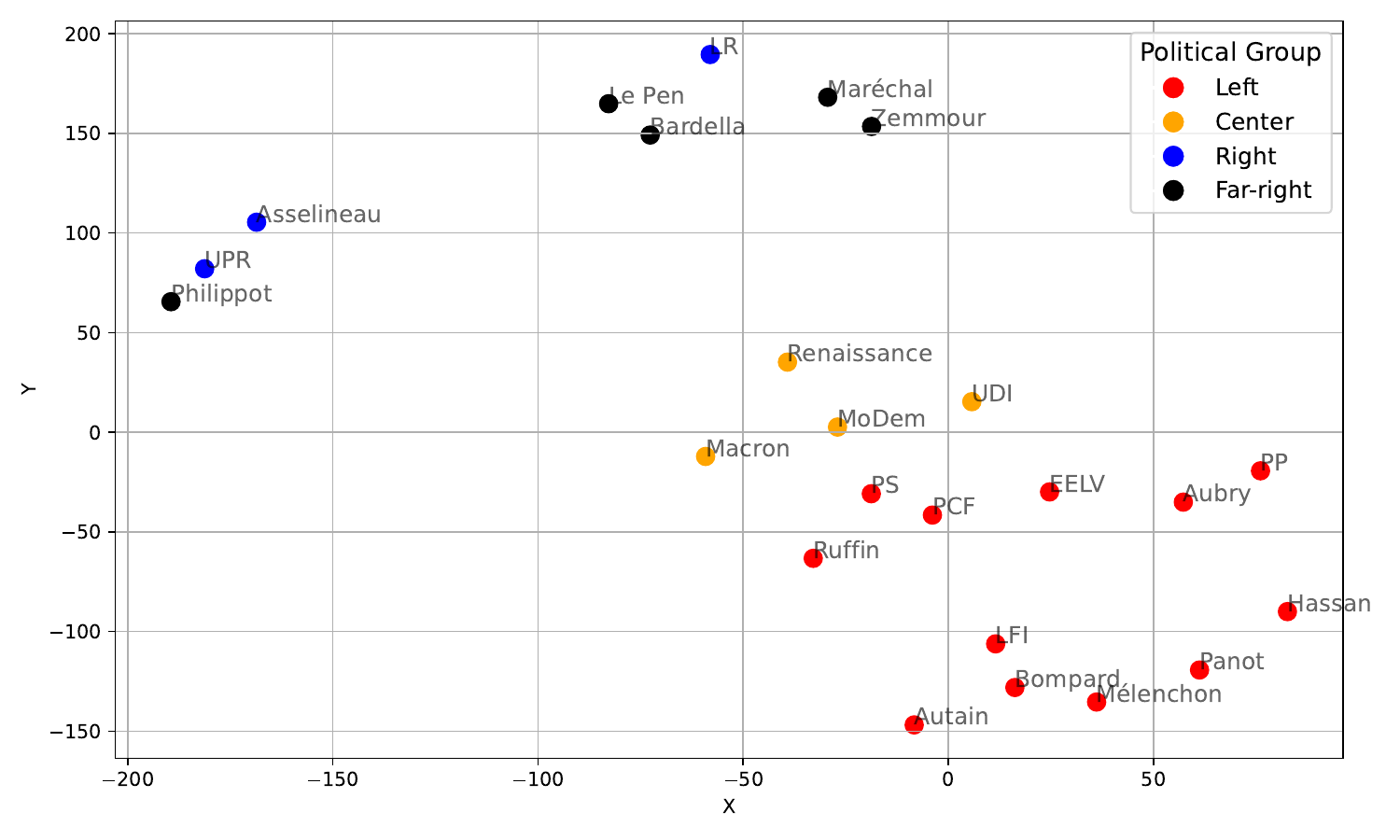}
  \caption{\textbf{Political dataset:} the t-SNE visualization of channel thematic coverage similarity.}
  \label{fig:polit_vis}
\end{figure}

\paragraph{Political dataset.} In the political dataset, the t-SNE plot shows even more apparent thematic clustering of French political figures based on their content (Figure \ref{fig:polit_vis}). The left-wing cluster can be divided into two subgroups. It includes the LFI party and its members, Mélenchon, Panot, Bompard, Hassan, and Autain, who have been in LFI for most of the data collection time and are positioned further from the center cluster. The second subgroup of left-wing parties and political figures, which includes parties like PS, PCF, and EELV, is located closer to the center, suggesting a more moderate thematic alignment. 

Center-leaning political figures form a coherent cluster between the left and right (including far-right) groups. However, their proximity to the left-wing cluster suggests a stronger thematic alignment with left-wing discourse.

The right and far-right clusters are clearly separated from the center and left clusters in the t-SNE plot, highlighting distinct thematic focus. These two clusters represent separate groupings, suggesting divergent narratives and issue priorities. Interestingly, the far-right cluster includes Les Républicains, a right-wing party, indicating thematic overlap with far-right discourse. One of the most frequent themes for channels in this cluster is \textit{Immigration \& Refugees}, highlighting a shared focus that likely contributes to their proximity in the t-SNE plot. The last cluster, on the left of the figure, includes figures like Philippot and Asselineau who have a strong popularity on YouTube (i.e., many subscribers) but are rather outsiders of the French political system attracting a very small percentage of the votes. Their key theme is \textit{News Media \& Freedom of Expression}, reflecting a recurring narrative that they are being censored or unfairly treated by mainstream media.

In summary, the visualization of the political dataset reflects current trends in the French political landscape. LFI appears further from the center than other left-wing parties, reflecting its more oppositional political strategy. Right-wing and far-right parties are positioned even more distinctly away from the center. This highlights the growing prominence and polarization of right-leaning narratives in contemporary political discourse.

\subsection{Stance Detection}

While thematic analysis shows what themes political actors engage with, it does not capture how they are portrayed. To address this, we created a labeled dataset of 65 transcripts indicating stance (Negative, Neutral, or Favorable) toward Macron, Bardella, and Mélenchon. We evaluated three Mistral models using accuracy and soft accuracy. Soft accuracy gives partial credit when Neutral is predicted instead of Favor or Against. Mistral Large performed best, with 74\% accuracy and 86\% soft accuracy. Our analysis focused on key parties, LFI (left), Renaissance (center), LR (right), RN (far-right), and their leaders, including Gabriel Attal and Jordan Bardella, reflecting their prominent roles in current discourse.

To select transcripts for stance detection, we used the RapidFuzz Python library \cite{fuzz}, which enables fuzzy matching, useful for handling misspellings and transcription errors common in auto-generated YouTube transcripts. Due to a low number of relevant transcripts for Ciotti (fewer than 50), we excluded him from the stance detection analysis. 

\begin{table}[h]
\centering
\begin{tabular}{lrrr}
\toprule
\textbf{Political Orient.} & \textbf{Against} & \textbf{Favor} & \textbf{Neutral} \\
\midrule
Center & 34.3\% & 2.6\%  & 63.1\% \\
Left   & 58.0\% & 0.2\%  & 41.8\% \\
Right  & 42.4\% & 3.0\%  & 54.7\% \\
\bottomrule
\end{tabular}
\caption{News media stances by political orientation toward all political figures.}
\label{tab:stance_distribution}
\end{table}

Table~\ref{tab:stance_distribution} presents a summary of how political figures and parties in general are portrayed by news media, grouped by the media's political orientation. Left-leaning outlets are highly critical, showing a strong tendency toward negative coverage with almost no favorable mentions. Right-leaning outlets are also largely critical but exhibit slightly more variation in tone, including a modest amount of favorable coverage. In contrast, center-aligned outlets are the most neutral, reflecting a preference for balance and journalistic restraint in their reporting.

\begin{table}[h]
\centering
\begin{tabular}{lrrrr}
\toprule
\textbf{Target} & \textbf{\# Videos} &\textbf{Against} & \textbf{Favor} & \textbf{Neutral} \\
\midrule
LFI & 1026&33.4\%& 1.3\% & 65.3 \%\\
Mélenchon &1102&33.8\% & 1.9\% & 64.3\% \\
\hline
Renaissance &1769 &35.3\%& 3.1\% & 61.6\% \\
Macron & 3670&45.6\%& 3.1\%& 51.3\% \\
Attal & 304&23.4\%& 4.9\% & 71.7 \%\\ 
\hline
LR& 131&15.3\% & 0.8\% & 84\% \\ 
\hline
RN& 2195&41.9\% & 2.5\%& 55.7\%\\
Le Pen & 2698&40.1\%& 2.3\% & 57.6\% \\
Bardella & 1510&51.6\%& 2.5\%& 45.9\% \\
\bottomrule
\end{tabular}
\caption{Distribution of stances toward different political figures across all news channels.}
\label{tab:stance_targets}
\end{table}

 Table~\ref{tab:stance_targets} shows that most political figures receive predominantly neutral coverage, with actors like Attal and Les Républicains exceeding 70\% neutral mentions. However, high-profile or polarizing figures, such as Bardella, Macron, and Le Pen, face significantly higher levels of negative coverage, with over 40\% of statements labeled as Against. Favorable mentions do not exceed 5\% across all targets. 
 
A closer look at the stance distribution by media orientation reveals further differences (Appendix Table \ref{tab:stance_percentages}). Bardella is the most criticized by both right- and center-leaning outlets, while Attal and Les Républicains receive the least negative coverage from the right. Among center-aligned media, Attal receives the most favorable and neutral treatment. In contrast, left-leaning media primarily target Macron, while Mélenchon and La France Insoumise (LFI) are less criticized. Overall, favorable coverage is rare, with neutral or critical reporting prevailing.

\section{Related Work}

\textit{Studies about YouTube.}
YouTube has become a central platform for political communication, with US users reportedly watching 100 times more candidate-related videos on YouTube than on CNN, C-SPAN, MSNBC, and Fox News combined~\cite{GOOGLE}. As a result, political video content has drawn increasing scholarly attention. Prior works have addressed ad classification challenges using multimodal approaches to handle tasks such as tone detection and distinguishing political ads from other online content~\cite{imabalnce_data, video_classification}.

In parallel, a growing body of work has examined harmful and extremist content on YouTube. \citet{rad_patways} analyzed over 330,000 videos and found that users often migrate toward more radical content over time. ~\citet{rec_system} explored the role of YouTube’s recommendation algorithm in amplifying far-right content, raising concerns about algorithmic bias and radicalization pathways.

YouTube's comment sections have also become a focus of research. \citet{jiang2019bias} examine the dynamics of comment moderation on YouTube, focusing on the perceived political bias in content removal. Their findings suggest that higher moderation rates are more strongly linked to ideological extremism and misinformation than to political orientation alone, challenging common assumptions about partisan censorship on the platform.  \citet{doi:10.1177/21582440221094587} proposed a user activity-based model to assess political polarization on South Korean YouTube channels, analyzing over 11 million comments from 600,000 users. Their study revealed a high degree of polarization, with only 8\% of users identified as neutral. \citet{wu2021cross} examined 134 million YouTube comments and found that while cross-partisan engagement exists, it is asymmetrical; conservatives are significantly more likely to engage with liberal content than liberals with conservative content.

Finally, ~\citet{10.1145/3614419.3644023} presented the first large-scale comparison of YouTube Shorts and regular videos, analyzing over 17 million videos across 70,000 channels. Their findings show that Shorts are popular among new creators and are mainly used for entertainment, while regular videos remain dominant in educational and political domains.


\textit{Studies about online news media.}
Many studies have examined online news, focusing on the dynamics of information flows ~\cite{RichardFletcher2020, doi:10.1126/sciadv.aau4586, Ognyanova2020, efstratiou2022adherence, papadamou2022just, mosleh2022measuring, Horne_Gruppi, 10.1145/3757540}. 
Complementing these works,~\citet{10646651} developed an automated method to identify and track misleading narratives across low-credibility news websites. Their analysis revealed that lesser-known outlets play a significant role in spreading misinformation and demonstrated how automated systems can support an early detection of emerging false narratives. 
\citet{chouaki} addressed limitations in platform access and introduced a data donation approach that captures users' interactions with news content. Their findings suggest that misinformation exposure is often driven more by users’ own content choices than by algorithms.


Contrary to these prior studies, we are the first to provide a large-scale, content-driven analysis of political discourse during the 2024 French elections by examining videos from three distinct groups of information sources on YouTube.
\change{\section{Limitations}}
\change{The video classification relied solely on transcripts, without incorporating visual cues that may provide important context, particularly in videos with very short transcripts. Most transcripts were automatically generated and therefore contain occasional spelling or recognition errors. While we did not find evidence of systematic bias in topic extraction, sporadic effects cannot be ruled out. Moreover, transcripts primarily reflect the narratives of content creators rather than the perspectives of their audiences, which limits insight into reception. Finally, the YouTube API does not provide data on dislikes or the temporal dynamics of engagement, restricting a more comprehensive understanding of audience sentiment and its evolution over time.
}

\section{Conclusion}

This paper examined French political and media discourse on YouTube before and during the politically intense period of the 2024 European and French legislative elections. 
By analyzing content from three types of actors, national news media, local news outlets, and political figures, we identified how different players emphasized distinct themes to shape the public debate. Our approach, combining large language models with clustering and manual review, allowed us to detect both prominent and context-specific issues. During our analysis, we identified themes that generated higher levels of engagement. Themes that generated the most engagement were often polarizing, with Antisemitism ranking highest for both center- and right-leaning news channels. In contrast, non-political themes like Video Games and Nature had the highest like-to-view ratios, suggesting that audience approval is higher for less controversial content. Finally, we examined how different news media outlets portrayed political figures. Overall, favorable coverage was uncommon across all media types, indicating a general tendency toward neutral or critical reporting. These findings show that both ideology and content style shape audience reactions, emphasizing YouTube’s evolving role in political communication.

\section{Acknowledgements}
We thank the anonymous reviewers for their valuable feedback, Vincent Vandersluis for proof-reading the paper, and the Hasler Foundation for funding this project.

\bibliography{aaai25}
\appendix
\section{Appendix}

\change{\paragraph{Prompt for topic extraction:}
For transcripts less than or equal to 500 words:
Prompt: You are given a transcript of the video. Detect one main topic from the given transcript. All topics should be no more than 3 words and in English.
Generate your response in the following way:\\
1. Topic1

For transcripts more than 500 words and less than or equal to 1000 words:
You are given a transcript of the video. Detect two main topics from the given transcript. All topics should be no more than 3 words and in English.
Generate your response in the following way:\\
1. Topic1\\
2. Topic2

For transcripts more than 1000 words and less than or equal to 1500 words:
You are given a transcript of the video. Detect three main topics from the given transcript. All topics should be no more than 3 words and in English.
Generate your response in the following way:\\
1. Topic1\\
2. Topic2\\
3. Topic3

For transcripts more than 1500 words and less than or equal to 2000 words:
You are given a transcript of the video. Detect four main topics from the given transcript. All topics should be no more than 3 words and in English.
Generate your response in the following way:\\
1. Topic1\\
2. Topic2\\
3. Topic3\\
4. Topic4

For transcripts more than 2000 words and less than or equal to 12000 words:
You are given a transcript of the video. Detect five main topics from the given transcript. All topics should be no more than 3 words and in English.
Generate your response in the following way:\\
1. Topic1\\
2. Topic2\\
3. Topic3\\
4. Topic4\\
5. Topic5

Larger transcripts were split into 1000-word parts and used the appropriate prompt.}

\begin{table*}[h!]
\small
    \centering
    \begin{tabular}{|p{9cm}|p{4cm}|p{4cm}|}
        \hline
        \textbf{Transcript} & \textbf{First Promt Output} & \textbf{Final Promt Output}\\ \hline
       Gonat, allez, la viande dans le sac, allez, voilà. Ah, chacun sa couette et du sommeil réparateur pour tous. C’est peut-être comme ça qu’on pourrait résumer la méthode de sommeil scandinave. Alors, ça paraît un peu bête comme ça, mais en fait, c’est assez logique, et c’est peut-être pour ça que cette technique venue du nord de l’Europe cartonne en ce moment sur TikTok. Sleple sov gka lugn F.
Ça évite, le soir, de dire : Tu prends ta couverture ? Ça tire de tous les côtés, quoi ! C’était la guerre, c’était la guerre. Donc, du coup, en Suède : Jag är så frusen på natten, så att jag blir inte så glad när någon tar mitt täcke. Chacun trouve son bonheur. Moi, ma femme est frileuse, donc elle, elle a vraiment la grosse couette. Moi, qui le suis beaucoup moins, je ne supporte pas d’avoir chaud au lit. Bah, moi, j’ai une couette légère, quoi. On dort mieux, on est moins stressé, on a une meilleure récupération musculaire.
C’est vrai que, moi, quand je suis en Suède, je dors beaucoup mieux. Là-bas, je me sens tout de suite à l’aise, quoi. Avec plus de confort avant de dormir. D’accord, il y a un petit truc quand on arrive chez des amis, on emmène donc une partie du voyage : Varför ska man inte ha ett eget täcke? Och därför tycker jag så jobbigt när man kommer till Frankrike att bara ha ett täcke. Dina föräldrar tycker det är väldigt konstigt och jobbigt med två täcken. De tänker att det här, varför ska man ha det så när vi är gifta? säger de iskallt.
Alors, viens, je vais te montrer ce qu’ils appellent le "stoua" ici, une chambre en plus, quoi. Voilà. Et là aussi, c’est pareil, il y a deux couettes. Partout, de toute façon, c’est comme ça.
  & Scandinavian Sleep Method, 
  Blankets/Coverlet, 
  Sleep Quality,
  Personal Preference, 
  Cultural Differences &Scandinavian Sleep Method\\
  \hline
Cet individu cagoulé s'apprête à dévaliser, en pleine journée, une bijouterie américaine avec l'aide, non pas de quelques complices, mais de plus d'une vingtaine de [Musique] braqueurs. Ce cambriolage géant s'est passé à Sunnyvale, dans la baie de San Francisco, au 791 East El Camino Real pour être précis. Après avoir détruit les portes à coups de marteau, les voleurs ont brisé les vitrines et récupéré leur butin en seulement 2 minutes, sans faire de [Musique] blessé.
Les suspects ont rapidement fui le magasin à bord de plusieurs véhicules. Arrivés sur les lieux, la police a identifié et poursuivi deux voitures, mais l'une d'entre elles a réussi à s'échapper. Au final, quatre individus ont été arrêtés avec leur butin en main, mais la valeur totale des objets volés reste inconnue. C'est le troisième braquage de ce type en moins d'un mois dans le quartier, visant en particulier des bijouteries gérées par des personnes d'origine indienne.
   & Jewelry Store Robbery, Large Scale Heist, Police Chase &Jewelry Store Robbery\\
   \hline
    \end{tabular}
    \caption{Examples of different prompts.}
    \label{tab:prompt_exmp}
\end{table*}

\begin{table}[!htbp]
\centering
\begin{tabular}{lcc}
\toprule
\textbf{Channel Name} & \textbf{No Transcr.} & \textbf{With Transcr.}\\ \\
\midrule
\multicolumn{3}{c}{\textbf{Left-Wing}} \\
\midrule
BLAST & 28    & 701   \\
Journal l'Humanité          & 12    & 478   \\
Le Média                    & 8     & 1105  \\
Mediapart                   & 13    & 405   \\
\midrule
\multicolumn{3}{c}{\textbf{Center}} \\
\midrule
28 minutes - ARTE         & 4     & 735   \\
AFP                       & 3802  & 2375  \\
ARTE                      & 249   & 1260  \\
C dans l'air              & 2001  & 36    \\
C à vous                  & 2499  & 1101  \\
FRANCE 24                 & 4716  & 12364 \\
France Inter              & 6964  & 155   \\
LCP  & 11    & 2113  \\
Le Monde                  & 277   & 469   \\
Le Nouvel Obs             & 1048  & 2     \\
LeHuffPost                & 2512  & 759   \\
Libération                & 13    & 365   \\
Marianne                  & 1     & 257   \\
Public Sénat              & 130   & 4164  \\
RFI                       & 2979  & 8     \\
RTL                       & 8486  & 0     \\
TF1 INFO                  & 88    & 3890  \\
TV5MONDE Info             & 306   & 11330 \\
euronews (en français)    & 2270  & 8640  \\
franceinfo                & 69    & 2436  \\
\midrule
\multicolumn{3}{c}{\textbf{Right-Wing}} \\
\midrule
BFMTV         & 1017  & 4806  \\
CNEWS         & 9     & 562   \\
Europe 1      & 36    & 21150 \\
LCI           & 27    & 3245  \\
L'Express     & 152   & 339   \\
Le Figaro     & 66    & 1188  \\
Le Parisien   & 1712  & 3687  \\
Le Point      & 3472  & 550   \\
Les Echos     & 359   & 50    \\
RMC           & 6     & 1894  \\
Sud Radio     & 4911  & 7615  \\
VA Plus       & 8     & 348   \\
\bottomrule
\end{tabular}%
\caption{\textbf{News Dataset}: Number of videos with and without transcripts, grouped by political orientation.}
\label{tab:len_counts_grouped}
\end{table}

\begin{table}[!htbp]
\centering
\begin{tabular}{lcc}
\toprule
\textbf{Channel Name} & \textbf{No Transcr.} & \textbf{With Transcr.}\\
\midrule
\multicolumn{3}{c}{\textbf{Left-Wing}} \\
\midrule
Clémentine Autain           & 0   & 57   \\
EELV   & 9   & 254  \\
François Ruffin             & 3   & 154  \\
Jean-Luc Mélenchon          & 9   & 300  \\
LFI         & 13  & 375  \\
Manon Aubry                 & 8   & 235  \\
Manuel Bompard              & 7   & 284  \\
Mathilde Panot              & 6   & 272  \\
PCF  & 34 & 546 \\
PRG       & 2   & 9    \\
PS       & 9   & 283  \\
Rima Hassan                 & 0   & 20   \\
PP              & 0   & 225 \\
\midrule
\multicolumn{3}{c}{\textbf{Center}} \\
\midrule
Emmanuel Macron             & 17  & 78   \\
Gabriel Attal               & 0   & 4    \\
MoDem                       & 20  & 476  \\
Renaissance                 & 2   & 28   \\
UDI                   & 1   & 107  \\
\midrule
\multicolumn{3}{c}{\textbf{Right-Wing}} \\
\midrule
Eric Ciotti                 & 2   & 12   \\
François Asselineau         & 0   & 94   \\
Nicolas Dupont-Aignan       & 326 & 4    \\
UPR  & 4  & 126  \\
les Républicains            & 2   & 26   \\
\midrule
\multicolumn{3}{c}{\textbf{Far-Right}} \\
\midrule
Florian Philippot           & 3   & 580  \\
Les Patriotes               & 0   & 16   \\ 
Jordan Bardella             & 20  & 170  \\
Marine Le Pen               & 18  & 133  \\
Marion Maréchal             & 6   & 153  \\
RN      & 244 & 0    \\
Éric Zemmour                & 2   & 104  \\
\bottomrule
\end{tabular}
\caption{\textbf{Political Dataset}: Number of videos with and without transcripts, grouped by political orientation.}
\label{tab:len_counts_grouped_polit}
\end{table}

\begin{table}[ht]
\centering
\begin{tabular}{l|r|r}
\hline
\textbf{Channel Name} & \textbf{No Transcr.} & \textbf{With Transcr.}\\
\hline
Charente Libre & 81 & 394 \\
Corse-Matin Presse & 168 & 485 \\
Groupe Nice-Matin & 121 & 582 \\
Journal La Montagne & 225 & 288 \\
La Nouvelle République  & 430 & 580 \\
La Provence & 1642 & 864 \\
La République des Pyrénées & 4 & 20 \\
La Voix du Nord & 32 & 150 \\
Le Dauphiné Libéré & 9 & 63 \\
Le Télégramme & 343 & 232 \\
Ouest-France & 531 & 919 \\
Presse Océan & 5 & 0 \\
Sud Ouest & 331 & 374 \\
\hline
\end{tabular}
\caption{\textbf{Local Dataset}: Number of videos with and without transcripts.}
\label{tab:local_subs}
\end{table}

\begin{table*}[h!]
\centering
\begin{tabular}{|l|l|r|r|r|}
\hline
\textbf{Political Orient.} & \textbf{Target} & \textbf{Against (\%)} & \textbf{Favor (\%)} & \textbf{Neutral (\%)} \\
\hline
\multirow{8}{*}{Left} & LFI & 23.5 & 0.0 & 76.5 \\
 & Mélenchon & 25.3 & 1.1 & 72.5 \\\cline{2-5}
& Renaissance & 56.5 & 0.5 & 43.5 \\
& Macron & 93.4 & 0.0 & 6.9 \\
 & Attal & 53.3 & 0.0 & 46.7 \\\cline{2-5}
 & Les Républicains & 29.3 & 0.0 & 70.7 \\\cline{2-5}
& Rassemblement National & 45.1 & 0.0 & 54.3 \\
& Le Pen & 46.8 & 0.4 & 53.6 \\
 & Bardella & 80.2 & 0.0 & 19.8 \\
\hline
\hline
\multirow{8}{*}{Center} & LFI & 26.3 & 1.1 & 69.6 \\
& Mélenchon & 26.7 & 0.8 & 71.3 \\
\cline{2-5}
 & Renaissance & 24.8 & 3.1 & 69.2 \\
 & Macron & 36.5 & 3.4 & 58.9 \\
& Attal & 19.0 & 8.0 & 72.0 \\
\cline{2-5}
& Les Républicains & 7.1 & 2.4 & 85.7 \\
\cline{2-5}
 & RN & 35.3 & 2.4 & 60.3 \\
 & Le Pen & 34.6 & 2.1 & 61.8 \\
 & Bardella & 46.0 & 1.9 & 49.5 \\
\hline
\hline
\multirow{8}{*}{Right}& LFI & 37.1 & 1.5 & 57.8 \\
 & Mélenchon & 38.6 & 2.6 & 59.1 \\
\cline{2-5}
& Renaissance & 36.6 & 3.4 & 57.3 \\
& Macron & 42.2 & 3.1 & 50.5 \\
& Attal & 22.2 & 3.5 & 70.2 \\
\cline{2-5}
& Les Républicains & 9.3 & 0.0 & 
90.7\\
\cline{2-5}
& Rassemblement National & 44.1 & 2.8 & 48.9 \\
 & Le Pen & 42.4 & 2.7 & 53.5 \\
 & Bardella & 49.6 & 3.3 & 45.5 \\
\hline
\end{tabular}
\caption{Distribution of stance by news media orientation and political figure.}
\label{tab:stance_percentages}
\end{table*}

\paragraph{Validation of figures 2 and 3:} We selected the clusters presented in Figure 2 for the news dataset and in Figure 3 for the political dataset. We represented each channel with a theme vector, based on how often its videos contained each theme (counting unique themes once per video). Afterward, we normalized these values to produce a probability distribution over themes in the same way it was done for the original figures.

To support the validity of the visual patterns observed in the plots, we quantitatively evaluated cluster quality using the following formula:
\[
Q_c = \frac{\overline{\cos(\mathbf{x}_i, \mathbf{x}_j)} \;\; i,j \in c}{\overline{\cos(\mathbf{x}_i, \mathbf{x}_k)} \;\; i \in c, k \notin c}
\]
where ${x}_i$ and ${x}_j$ are theme vectors of the channels in the same cluster, $c$, and ${x}_k$ denotes vectors of channels from different clusters.
The numerator represents the average cosine similarity between items within the same cluster, and the denominator represents the average cosine similarity between items in the cluster and items outside it.

For the news dataset, the cluster containing France 24, TV5Monde Info, Euronews, and LCI achieves the highest quality score (1.4), indicating that these outlets form a particularly tight and distinct group. This is followed by the cluster of left-leaning media (Blast, Le Média, Mediapart, Journal l’Humanité) and VA Plus, and another cluster including AFP, Le Parisien, BFMTV, and CNEWS, both with scores around 1.3, suggesting strong internal coherence but some overlap with other groups. By contrast, the final cluster, including outlets such as Le Monde, Marianne, and Le Figaro, yields the lowest score (1.01), showing weaker distinctiveness compared to other clusters. These results are consistent with the clustering structures observed in Figure 2.

For the political dataset, clusters centered on Asselineau, Philippot, and UPR (1.3), as well as those including Bompard, LFI,  Panot,  Mélenchon (1.2), and Maréchal, Zemmour, and LR (1.18), all display strong internal cohesion and clear group structure. By contrast, the cluster bringing together MoDem, PCF, Renaissance, and Place Publique (from left and center political groups) records the lowest score (1.13), reflecting weaker internal consistency. These findings align with the cluster structures shown in Figure 3.

\end{document}